# Photonic analogue of quantum spin Hall effect

Cheng He[1]*, Xiao-Chen Sun[1]*, Xiao-ping Liu[1], Ming-Hui Lu[1†], Yulin Chen[2], Liang Feng[3] and Yan-Feng Chen[1†]

[1]*National Laboratory of Solid State Microstructures & Department of Materials Science and Engineering, Nanjing University, Nanjing 210093, China*

[2]*Clarendon Laboratory, Department of Physics, University of Oxford, Parks Road, Oxford, OX1 3PU, UK*

[3]*Department of Electrical Engineering, University at Buffalo, The State University of New York, Buffalo, NY 14260, USA*

*These authors contributed equally to this work.

†Correspondence and request for materials should be addressed to M. H. Lu (luminghui@nju.edu.cn) and Y. F. Chen (yfchen@nju.edu.cn).

**Symmetry-protected photonic topological insulator exhibiting robust pseudo-spin-dependent transportation[1-5], analogous to quantum spin Hall (QSH) phases[6,7] and topological insulators[8,9], are of great importance in fundamental physics. Such transportation robustness is protected by time-reversal symmetry. Since electrons (fermion) and photons (boson) obey different statistics rules and associate with different time-reversal operators (i.e., $T_f$ and $T_b$, respectively), whether photonic counterpart of Kramers degeneracy is topologically protected by bosonic $T_b$ remains unidentified. Here, we construct the degenerate gapless edge states of two photonic pseudo-spins (left/right circular polarizations)**

**in the band gap of a two-dimensional photonic crystal with strong magneto-electric coupling. We further demonstrated that the topological edge states are in fact protected by $T_f$ rather than commonly believed $T_b$ and their pseudo-spin dependent transportation is robust against $T_f$ invariant impurities, discovering for the first time the topological nature of photons. Our results will pave a way towards novel photonic topological insulators and revolutionize our understandings in topological physics of fundamental particles.**

Recent topological description of electronic quantum behaviors sets in a new paradigm in the classification of condensed matters, including $T_f$ broken integer quantum Hall effect due to the external magnetic bias, $T_f$ invariant $Z_2$ quantum spin Hall (QSH)[10,11] and topological insulators[12,13] originated from the intrinsic strong spin-orbit coupling. Besides, with time-periodic oscillated electron potential, Floquet topological insulators show topological protection under very weak or even without spin–orbit coupling[14]. A remarkable topological property is the robust charge or spin-dependent transportation against impurities and perturbations as long as symmetry is intact. In the realm of optics and photonics, realizing the non-trivial photonic topological phase can revolutionize our understandings on optics and photonics as well as the emergent artificial photonic structural materials (e.g. dielectric superlattice[15], photonic crystals[16,17] and metamaterials[18]) and thus fundamentally inspire new ways to manipulate light and its propagation and scattering properties. Recently there has been rapid progress in searching for photonic topological states. The photonic analogue of the integer quantum Hall effect was theoretically and experimentally demonstrated using single TE[19,20] or TM[21,22] (transverse electric or magnetic) polarization in gyrotropic photonic crystals with broken time-reversal symmetry (*T*) under an external magnetic bias. Pseudo-spin-dependent photonic topological states were also achieved utilizing a pair of degenerate photonic states with different spin-orbit interactions, i.e., different states experience opposite effective magnetic gauge fields[2,23,24]. More recently, to eliminate the limitations of external magnetic bias, hybrid TE+TM/TE-TM[1,4,5] states, clockwise/counter-clockwise[25,26] states, and photonic Floquet topological insulators[3] have been explored to realize photonic QSH effects under the time-reversal symmetry invariant. All these photonic topological states enable a flexible platform to investigatefundamental physics of time-reversal symmetry of particles with different spin quantum numbers.

Despite similarities between photons and electrons, they obey different statistics rules and possess different spin quantum numbers. Consequently photons and electrons have different physical forms of the time-reversal operator[27]: $T_b$ for photons and $T_f$ for electrons, respectively, defined as $T_b = \tau_z K$ where $\tau_z = diag\{1, -1\}$ is the $z$-component of Pauli matrix and $K$ is complex conjugation, while $T_f = \tau_z K \tau_x = i\tau_y K$ where $\tau_y = [0, -i; i, 0]$ is the $y$-component of Pauli matrix. In this letter, we will start from the above fundamental symmetries of photons and electrons to investigate the topological nature of photons and the intrinsic protection mechanism of photonic topological Kramers degeneracy. We will exploit left- and right-circular polarization states as a pair of degenerate photonic pseudo-spin states to realize the Kramers degeneracy through strong magneto-electric coupling in a judiciously designedphotonic crystal consisting of piezoelectric (PE) and piezomagnetic (PM) superlattice constituents. Associated with the achieved Kramers degeneracy in our photonic crystal, photonics QSH effects are well demonstrated, with a bulk band gap and a pair of gapless edge states for lef and right circular polarization states respectively, which show robust one-way pseudo-spin-dependent propagation against impurities if the system is invariant under $T_f$ operation. However,such one-way propagation is sensitive to impurities with their optical parameters satisfying $T_b$ invariant but $T_f$ breaking. With the rigorous analysis of the fundamental symmetry in our system, it is evident that the robustness of photonic topological states is in fact protected by $T_f$ rather than intuitively believed $T_b$.

Analogous to electronic spin-up and spin-down, a pair of degenerate left-circular polarization(LCP: positive helix or pseudo-spin up) and right-circular polarization (RCP: negative helix or pseudo-spin down) photonic states(Fig. 1a) can be described using Beltrami fields[28]

$$\begin{bmatrix} \psi_{LCP} \\ \psi_{RCP} \end{bmatrix} = \begin{bmatrix} 1 & i \\ i & 1 \end{bmatrix} \begin{bmatrix} E_z \\ H_z \end{bmatrix}. \tag{1}$$

The Kramers degeneracy is critical to create impurity-insensitive topological states, however, typical LCP and RCP polarization states do not satisfy the Kramers degeneracy under the $T_b$ operation: $T_b \psi_{LCP} = \psi_{LCP}^*$ and $T_b \psi_{RCP} = -\psi_{RCP}^*$. Here, we exploit a different symmetry operator by exploring the magneto-electric coupling and its associated interchange operator ($\tau_x = [0,1;1,0]$, the *x*-component of Pauli matrix) $E_z \leftrightarrow H_z$ to fulfill the requirement of Kramers degeneracy for LCP and RCP states. By combining $T_b$ operator and interchange operator $\tau_x$, a new symmetry operator in this model can be represented as $T_f = \tau_z K \tau_x = i\tau_y K$ ($\tau_y = [0,-i;i,0]$ is the *y*-component of Pauli matrix) that is interestingly the same as the time-reversal symmetry operator of fermions. In other words, the photonic Kramers degeneracy can be created if the delicate magnetic-electric coupling is introduced. Therefore, the new operator considering magneto-electric coupling in our system allows us to directly treat LCP and RCP states as a pair of pseudo-spin states satisfying the Kramers degeneracy $T_f \psi_{LCP} = \psi_{RCP}^*$ and $T_f \psi_{RCP} = -\psi_{LCP}^*$.

To realize pseudo-spin-dependent light propagation for mimicking electronic QSH, we studied a quasi-two-dimensional photonic crystal consisting of arrayed meta-atoms: a PE and PM superlattice with its periodicity in a deep subwavelength scale along the *z* direction (Fig. 1b).We utilized the strong magneto-electric coupling inherently associated with the PE and PM superlattice[29] to realize the pseudo-spin-orbit coupling and Kramers degeneracy for LCP and RCP. In our system, the Hamiltonian for LCP and RCP states is,

$$\mathcal{H}\begin{bmatrix}\psi_{LCP}\\ \psi_{RCP}\end{bmatrix}=\begin{bmatrix}\mathcal{H}_0-i\mathcal{H}_1 & 0\\ 0 & \mathcal{H}_0+i\mathcal{H}_1\end{bmatrix}\begin{bmatrix}\psi_{LCP}\\ \psi_{RCP}\end{bmatrix}=0, \qquad (2)$$

where $\mathcal{H}_0=k_0^2\varepsilon_{zz}+\partial_x\frac{1}{\mu_{//}}\partial_x+\partial_y\frac{1}{\mu_{//}}\partial_y$, $\mathcal{H}_1=\partial_x\kappa\partial_y-\partial_y\kappa\partial_x$, $\mu_{//}=(\mu_{xx}\varepsilon_{xx}-\xi_{xy}^2)/\varepsilon_{xx}$, and $\kappa=-\xi_{xy}/(\mu_{xx}\varepsilon_{xx}-\xi_{xy}^2)$ is the pseudo-spin-orbit coupling parameter, showing some polariton dispersion behavior near the resonant frequency $\omega_T$ (Fig. 1c).The $\pm i\mathcal{H}_1$ operator indicates the pseudo-spin-orbit coupling and the resulted opposite effective gauge fields for LCP and RCP states. For example, in our system, the parameters at the frequency of $1.076\omega_T$: $\kappa=0.3$, $\mu_{//}=\varepsilon_{//}=2.5$ and $\mu_{zz}=\varepsilon_{zz}=14.2$ (see Methods and Supplementary Materials SA for details).

The band structure of the photonic crystal (Fig. 2a) clearly shows that the second and third bulk bands cross at **M** point of the Brillouin zone without pseudo-spin-orbit coupling effect, *i.e.*, $\kappa=0$; while if pseudo-spin-orbit coupling is introduced in the system, *e.g.*, $\kappa=0.3$, this degeneracy would be lifted, leading to a bandgap. In the bandgap of the bulk states, however, there exist gapless helical edge states, corresponding to two pseudo-spin states of LCP and RCP polarizations (Fig.2b). The dispersion curves of two edge states cross with each other, forming a Dirac cone with a four-fold degeneracy at the **M** point and pseudo-spin-dependent one-way edge-state propagation in our system (Figs. 2c-2f). Since LCP and RCP states experience opposite effective gauge fields, their power fluxes on the edge of the photonic crystal correspond to different clockwise and counter-clockwise fashions, respectively, and the resulted one-way propagations are thus opposite with each other.

To illustrate the robustness of one-way transportation, we studied the backscattering immune propagations under different cases (Fig. 3). Without any defects, the pseudo-spin-dependent transportation is validated as LCP (RCP) source only excites one-way clockwise (anti-clockwise) propagation of LCP (RCP) light (Fig. 3a) (see Supplementary Materials SB for details). The demonstrated one-way robust transportation is further verified against three different types of defects as shown in Figs. 3b-3d, including an L-shape slab obstacle, a cavity obstacle and a strongly disordered domain inside the cavity, respectively. It is evident that the helical edge states propagate along their preferential directions without inducing any backscattering. Note that this kind of transportation behavior is in fact consistent with the observation in Ref. [1].

In addition, the robustness of such photonic pseudo-spin-dependent propagation is further investigated against perturbations from all four possible types of impurities with their optical parameters satisfying different $T_b$ and $T_f$ symmetry: (i), a uniaxial dielectric impurity which keeps $T_b$ invariant but breaks $T_f$; (ii), a Tellegen impurity which breaks both $T_b$ and $T_f$; (iii), a chiral impurity which maintains $T_b$ invariant and breaks $T_f$; and (iv), a chiral impurity which keeps both $T_b$ and $T_f$ invariant. In case (i) and (ii), unfortunately, LCP light is backscattered into RCP light by the impurities (left panel of Figs. 4a and 4b), suggesting that regardless of the impurity's $T_b$ symmetry the pseudo-spin dependent propagation will not be robust as long as $T_f$ symmetry is broken. In other words, $T_b$ invariant proposed in the past[1] cannot guarantee the robust pseudo-spin-dependent transportation. This conclusion is consistent with the evolution of the projected band structures (right panel of Figs. 4a and 4b) from a gapless state to a current state with a band gap. In case (iii), the eigen equation of such chiral medium can be casted into two independent LCP and RCP eigen states with different Bloch wave vectors. Although the impurity cannot backscatter RCP light (left panel of Fig. 4c), the LCP and RCP edge states are decoupled and (right panel of Fig. 4c) exhibit two independent integer photonic quantum Hall effect, which cannot longer be consider as an analogue of QSH. However, incase (iv), RCP light is excited at the impurity site, but only located in a nearby region as shown in left panel of Fig. 4d. Remarkably, no RCP light propagates back. It is because such $T_f$ invariant impurity is a polarization degenerate medium for TE+TM and TE-TM states. With an LCP light incidence, the light inside the rod is decoupled into such two degenerate eigen states, which would be finally restored to LCP light when propagating outside. Similar to electronic QSH effect, destructive interference between two backscattering paths leads to backscattering immune and robust one-way transmission when scattered by a $T_f$ invariant impurity. This unique transportation characteristic can be confirmed by examining the corresponding projected band structure (right panel of Fig. 4d), in which the edge states remains gapless and continues to support pseudo-spin-dependent transportation. It is therefore evident that LCP and RCP eigenstates can only be treated as a pair of pseudo-spin states under the $T_f$ operator, $i.e., T_f \psi_{LCP} = \psi^*_{RCP}$ and $T_f \psi_{RCP} = -\psi^*_{LCP}$ and that the eigen equation (2) of our system only satisfies $T_f \mathcal{H} T_f^{-1} = \mathcal{H}$. We should also emphasize that the Hamiltonian in our system does not satisfy $T_b \mathcal{H} T_b^{-1} \neq \mathcal{H}$. Therefore, photonic QSH in our proposed system is protected by $T_f$ rather than commonly believed $T_b$, satisfying $T_f^2 = (i\tau_y K)^2 = -1$.

Our observation of $T_f$ symmetry protected photonic topological phase is by no means limited to only the case we consider in this letter. In fact it can be applied to the previously reported photonic topological insulator systems[1,4,5], for example: the hybrid TE+TM and TE-TM topological states discussed in Ref. [1] can be represented using a pair of pseudo-spin states under the $T_f$ operator, *i.e.*, $T_f(E_z + H_z) = (E_z - H_z)^*$ and $T_f(E_z - H_z) = -(E_z + H_z)^*$, and the corresponding eigen equations are $T_f$ invariant as well. The backscattering behavior of such hybrid eigen states against four types of impurities is similar to that of our system (see Supplementary Materials SC for details).

In summary, we construct a $T_f$ invariant Tellgen magnetic-electric coupling photonic crystal system consisting of stacked PE and PM superlattice constituents embedded in vacuum background. Analogue to 2D topological insulators, a photonic counterpart of QSH effect are well demonstrated in such a system, where the pseudo-spin states are represented by LCP and RCP edge states and they exhibit one-way pseudo-spin-dependent propagation. More importantly, it is the first time evident that that the robustness of this unique transportation property is protected by $T_f$ rather than commonly believed $T_b$ symmetry, which is in general applicable to all previously studied photonic topological insulator systems,[1-5,25,26] confirmed with detailed numerical analysis. Our findings are of great importance to revolutionize our understanding in topological nature of fundamental particles and may pave the way towards a new class of photonic topological states.

## Methods

**Effective bi-anisotropic tensor representation:**

To obtain the effective constitutive relation of PE and PM superlattice, we resort to classical PE and PM equations dealing with coupling effect between the EM waves and lattice vibrations:

$$T_I = C_{IJ}S_J - e_{Ij}(z)E_j - m_{Ij}(z)H_j,$$
$$D_i = e_{iJ}(z)S_J + \varepsilon_0\varepsilon_{ij}E_j, \qquad (i,j=1,2,3; I,J=1,2,...6)$$
$$B_i = m_{iJ}(z)S_J + \mu_0\mu_{ij}H_j. \tag{3}$$

$T_I$, $S_J$, $E_j(H_i)$, $D_i$ ($B_i$) and $\varepsilon_{i,j}$ ($\mu_{i,j}$) are the stress, strain, electric field (magnetic field), electric displacement (magnetic flux density) and permittivity (permeability) respectively. The elastic coefficient is $C_{IJ}^E = C_{IJ} - i\omega\gamma$ with the damping coefficiency $\gamma$. The periodically modulated PE (PM) coefficients are $e_{iJ}(z) = e_{iJ}f(z)$ [$m_{iJ}(z) = m_{iJ}g(z)$], where the modulation functions is $f(z) = 0,1$ [$g(z) =1, 0$] in the corresponding domains. $i(j)$ =1, 2, and 3 representing the $x$-, $y$- and $z$-axis respectively, $I(J)$ =1, 2, 3, 4, 5, and 6, indicating the abbreviation subscript of the tensor. (see Supplementary Materials SA for details). Here, the point groups of PE and PM materials we use are *622* and *6mm* (or *422* and *4mm*) respectively, with nonzero PE coefficient $e_{14}$ and PM coefficient $m_{15}$. As the EM wavelength is much larger than the period of superlattice, the effective bi-anisotropic constitutive relation can be described as

$$\boldsymbol{D} = \hat{\varepsilon}\boldsymbol{E} + \hat{\xi}\boldsymbol{H},$$
$$\boldsymbol{B} = \hat{\varsigma}\boldsymbol{E} + \hat{\mu}\boldsymbol{H}. \tag{4}$$

Here, $\hat{\varepsilon} = diag\{\varepsilon_{xx}, \varepsilon_{xx}, \varepsilon_{zz}\}$, $\hat{\mu} = diag\{\mu_{xx}, \mu_{xx}, \mu_{zz}\}$, $\hat{\xi} = \hat{\varsigma}^T$, $\xi_{yx} = \xi_{xy} = A_1 e_{14} m_{15}$, $A_1 = 2/[d^2\rho(\omega^2 - \omega_T^2 + i\gamma_T\omega)]$, and $\omega_T = G_1\sqrt{C_{44}/\rho}$ representing transverse vibration frequency. In lossless condition ($\gamma_T = 0$), $\hat{\xi} = \hat{\varsigma}^\dagger$ (superscript dagger represents complex conjugate transpose) and it only has real tensor elements leading to the so called Tellegen media[30], whose magnetoelectric effect breaks the $T_b$ symmetry and reciprocity ($\hat{\xi} \neq -\hat{\varsigma}^T$)[31]. Fig. 2 is calculated based on equations (4). Numerical investigations in this letter are conducted using hybrid RF mode of commercial FEM software (COMSOL MULTIPHYSICS 3.3).

**Theoretical model.** In order to establish an analytical model via low-energy effective Hamiltonian,

TBA approach is used to analyze our system where the LCP can RCP states can be treated independently. For the LCP mode there are three types of eigen states, $|s^{LCP}\rangle$, $|p_x^{LCP}\rangle$ and $|p_y^{LCP}\rangle$. The mode hopping behavior among these eigen-modes is analyzed with TBA method and summarized as follows: The interaction between $|p_x^{LCP}\rangle$ and $|p_y^{LCP}\rangle$ only contains the on-site coupling, while the interaction between other pairs contains both the on-site coupling and the nearest-neighboring coupling. Herein, we define new $p$ states as the linear superposition of $|p_x^{LCP}\rangle$ and $|p_y^{LCP}\rangle$ with $|p_\pm^{LCP}\rangle = \mp\frac{1}{\sqrt{2}}\left(|p_x^{LCP}\rangle \pm i|p_y^{LCP}\rangle\right)$. By applying the perturbation theory and expanding the $\boldsymbol{k}$ dependent Hamiltonian to the first order, we can obtain the low energy approximated effective Hamiltonian as following,

$$H_{LCP} = \sigma_0 + m_0\tau_z + \hbar v_f(k_y\tau_x + k_x\tau_y) \to H_0, \qquad (5)$$

where $\sigma_0$ is the energy origin point, $m_0$ is mass term and $v_f$ represents the phase velocity near the Dirac point. It can also be written in the form of second quantization:

$$H_0(\vec{k}) = \epsilon_0\left(c_s^\dagger c_s + c_p^\dagger c_p\right) + m_0\left(c_s^\dagger c_s - c_p^\dagger c_p\right) + \hbar v_f k_y\left(c_s^\dagger c_p + c_p^\dagger c_s\right) + i\hbar v_f k_y\left(-c_s^\dagger c_p + c_p^\dagger c_s\right). \qquad (6)$$

Clearly the Hamiltonian here has a Dirac form. The Hamiltonian of LCP state can therefore be expressed in terms of the RCP state via symmetry operation on all eigen states. The total Hamiltonian can then be written as

$$H = \begin{bmatrix} H_0(\boldsymbol{k}) & 0 \\ 0 & H_0^*(-\boldsymbol{k}) \end{bmatrix}. \qquad (7)$$

This total Hamiltonian representation can exactly map to the Hamiltonian in BHZ model of 2D electronic QSH effect[7], which remains invariant under $T_f$ operation. By solving the total Hamiltonian, the dispersion of the edge states for our system can be determined as $E = \sigma\hbar v_f k_y, \sigma = \pm 1$. This equation can also be used to calculate $Z_2$ invariant. Due to the inversion symmetry of the model[32]. The $Z_2$ invariant can be determined by the quantity equation, $\delta_i = \prod_{m=1}^{N} \xi_{2m}(\Gamma_i)$. Here, $\xi_{2m}(\Gamma_i)$ is the parity eigenvalue of the 2m$^{th}$ occupied energy band at four $T_f$ invariant momenta $\Gamma_i$ in the Brillouin zone. The $Z_2$ invariant $v = 0,1$, which distinguishes the quantum spin-Hall phase, is governed by the product of all the $\delta_i$: $(-1)^v = \prod_i \delta_i$. As for the $4 \times 4$ matrix discussed in equation (7), the quantity equation can be further simplified as $\delta_i = \xi(\Gamma_i)$, in which $\xi(\Gamma_i)$ is the parity eigenvalue of lower energy band at $\Gamma_i$. Because of $T_f$ invariance, the eigen states have degeneracy at $\Gamma_i$. We can find that the Bloch field distribution at (0, 0) point has even parity with parity eigenvalue 1, while the Bloch field distributions at other three points have odd parity with parity eigenvalues -1. Therefore, $(-1)^v = -1$ and the $Z_2$ invariant is $v = 1$, which imply the non-trivial topological states. (See Supplementary Materials SD for details).

**Acknowledgements**

The work was jointly supported by the National Basic Research Program of China (Grant No. 2012CB921503 and No. 2013CB632702) and the National Nature Science Foundation of China (Grant No. 1134006). We also acknowledge the support from Academic Program Development of Jiangsu Higher Education (PAPD) and China Postdoctoral Science Foundation (Grant No. 2012M511249 and No. 2013T60521). Y.L. Chen acknowledges support from a DARPA MESO project (No. 187 N66001-11-1-4105).

**Author contributions**

C. H. and M. -H. L. conceived the idea, C. H. performed the numerical calculation. X.-C.S.,C. H.

and M.-H. L. carried out the theoretical derivation and analysis. X. -P. L., Y. L. C. and L. F. assisted in analyzing the results. All the authors contributed to discussion of the project and manuscript preparation. M.-H.L. and Y.-F.C. supervised all of the work and guided this project.

**Figure captions：**

**Figure 1 | Schematics of the photonic QSH system represented by LCP and RCP states .a**, Degenerate LCP and RCP photonic states (pseudo-spin up and pseudo-spin down) analogous to spin up and spin down of electrons. Note that we have set the impedance $\eta = \sqrt{\mu/\varepsilon} = 1$ ($\mu/\varepsilon$ corresponds to permeability/permittivity) for simplicity. **b**, Two-dimensional square-lattice photonic crystal consisting of stacked PE and PM superlattice constituents in vacuum background. The positive domain and stacking direction of PE (red arrows) and PM (black arrows) media are along z-axis. Lattice constants of square photonic crystal lattice and PE/PM superlattice are *a* and *2d*, respectively. The radius of cylinder is *r*=0.11*a*. LCP and RCP edge states propagate in xy-plane at the boundary (with a separation distance of 0.5*a*) between square-lattice photonic crystal and cladding layer ($\varepsilon = \mu = diag\{1,1,-1\}$). **c,** Dispersive pseudo-spin-orbit coupling parameter $\kappa$ defined in equation (2). In this plot, open and solid circles represent the real and imaginary parts, respectively. **d,** Zoom-in view of the rectangular frequency area around frequency of interest. In our structure, we consider a loss $\gamma_T = 0.001$, and parameter $\kappa = 0.3$ at the frequency of $1.076\omega_T$ (black arrow), which is away from the polariton frequency $\omega_T$.

**Figure 2| Band structures and Bloch field distributions**. **a,** Bulk band structures of the photonic crystal without pseudo-spin coupling effect (dotted lines, $\kappa = 0$) and with pseudo-spin coupling effect (solid lines, $\kappa = 0.3$). **b,** The projected band structures with $\kappa = 0.3$, where bulk states are denoted by blue lines and gapless LCP and RCP edge states are denoted by green and red lines, respectively. **c-f,** The Bloch field distributions of supercell configuration corresponding to points **c-f** marked in Fig. 3b, respectively. Lower panels are the zoom-in view of the rectangular region, in which opposite rotating power flux near the boundary is represented by black arrows. Blue and red colors represent negative and positive field values. To keep the same unit between LCP and RCP, we set RCP $E_z - iH_z$ instead of $iE_z + H_z$ in equation (1) in field simulations.

**Figure 3| Robustness of one-way pseudo-spin-dependent propagation**. **a,** Backscattering immune pseudo-spin-dependent propagation of LCP (upper panel) and RCP (lower panel) light.

**b-d,** Robustness of the helical edge states for LCP (upper panel) and RCP (lower panel) light against different types of defect: (**b**) a L-shape slab obstacle (index of refraction -$i$) with thickness $0.5a$, (**c**) a cavity obstacle and (**d**) a strongly disordered domain in the cavity. Operating frequency is $0.6(2\pi c/a)$. LCP and RCP point excitation sources are indicated by green and red stars, respectively. Blue and red colors represent negative and positive field values.

**Figure 4| $T_f$ symmetry-protected propagation against four types of impurities**. **a,** dielectric impurity ($\boldsymbol{\varepsilon} = diag\{1,1,4\}$, $\boldsymbol{\mu} = \boldsymbol{I}$), **b,** Tellegen ($\boldsymbol{\varepsilon} = \boldsymbol{\mu} = \boldsymbol{I}$, $\xi_{zz} = \varsigma_{zz} = 1$) impurity, **c**, chiral impurity ($\boldsymbol{\varepsilon} = \boldsymbol{\mu} = \boldsymbol{I}$, $\xi_{zz} = -\varsigma_{zz} = i$), **d**, chiral impurity ($\xi_{xy} = -\xi_{yx} = 0.3i$, only possessing pure imaginary parts, $\boldsymbol{\zeta} = -\boldsymbol{\xi}^T$, other parameters are the same as superlattice) which remains invariant under $T_f$ operation. In the left panel of above four cases, a LCP point excitation source (green stars) at a frequency of $0.6(2\pi c/a)$ is used to study the pseudo-spin-dependent propagation property, and cylindrical impurity with radius $r = 0.11a$ is placed near the boundary. Colorbar represents the field's absolute value. Corresponding projected band structures are plotted in the right panel.

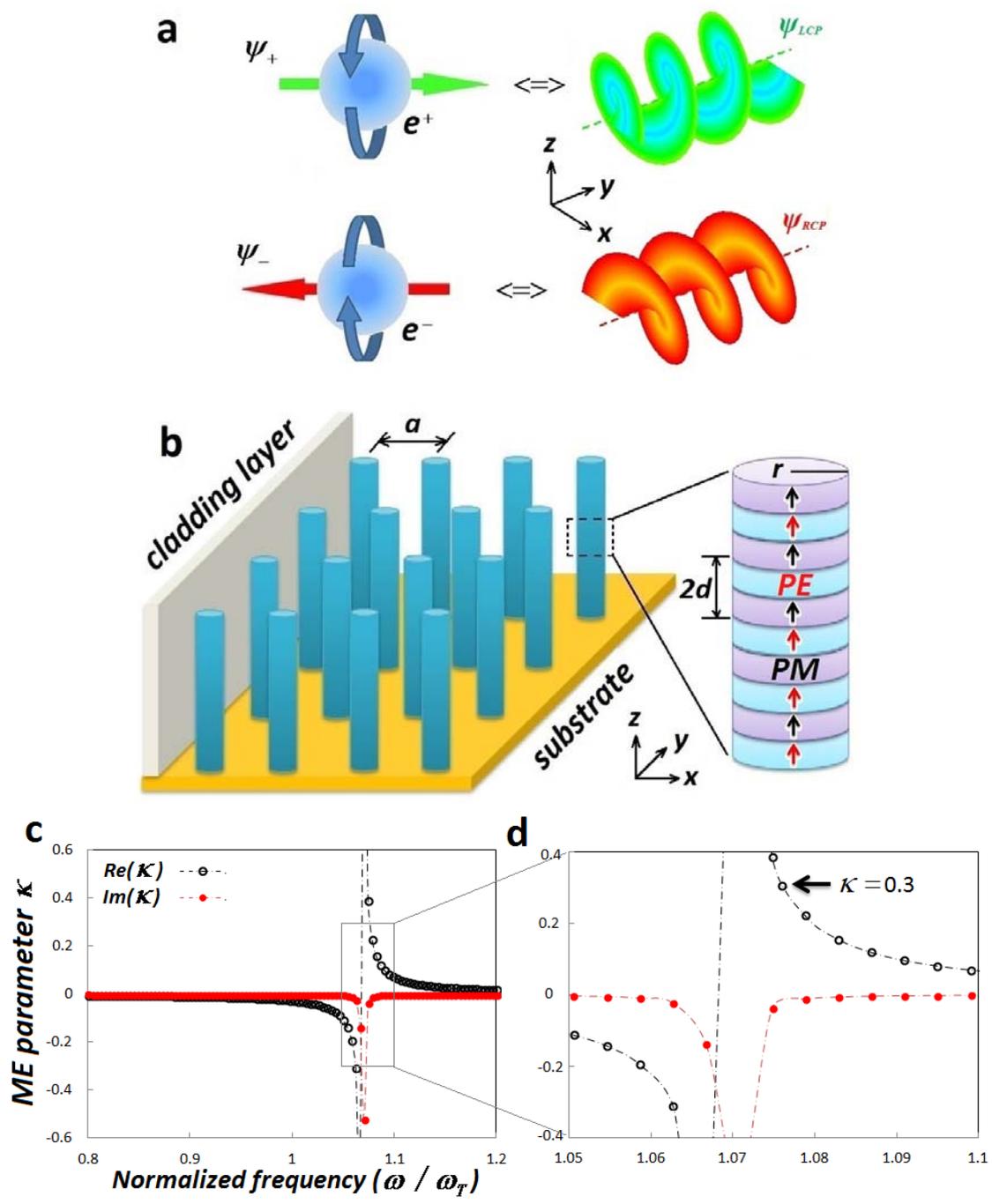

**Figure 1**

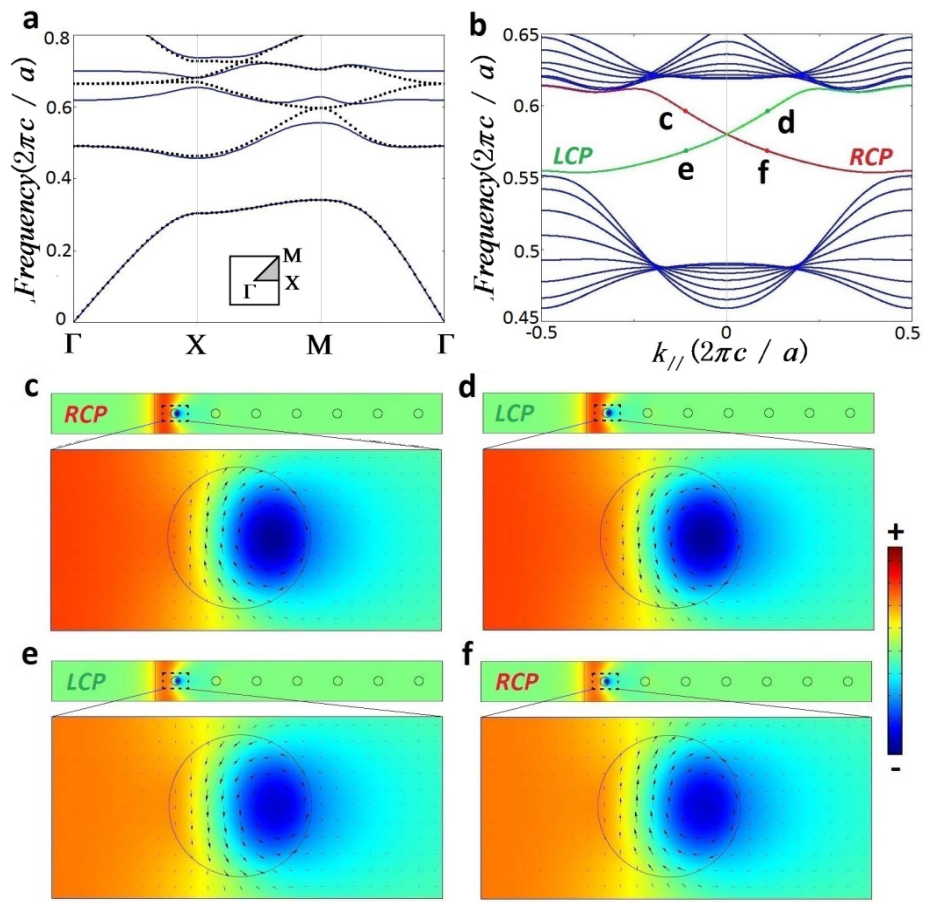

**Figure 2**

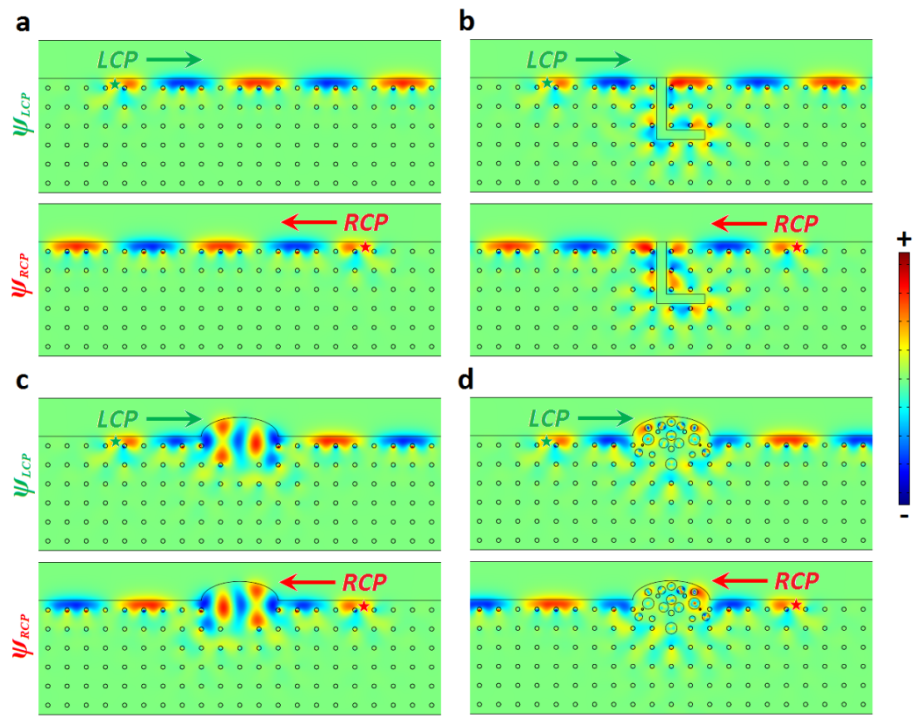

**Figure 3**

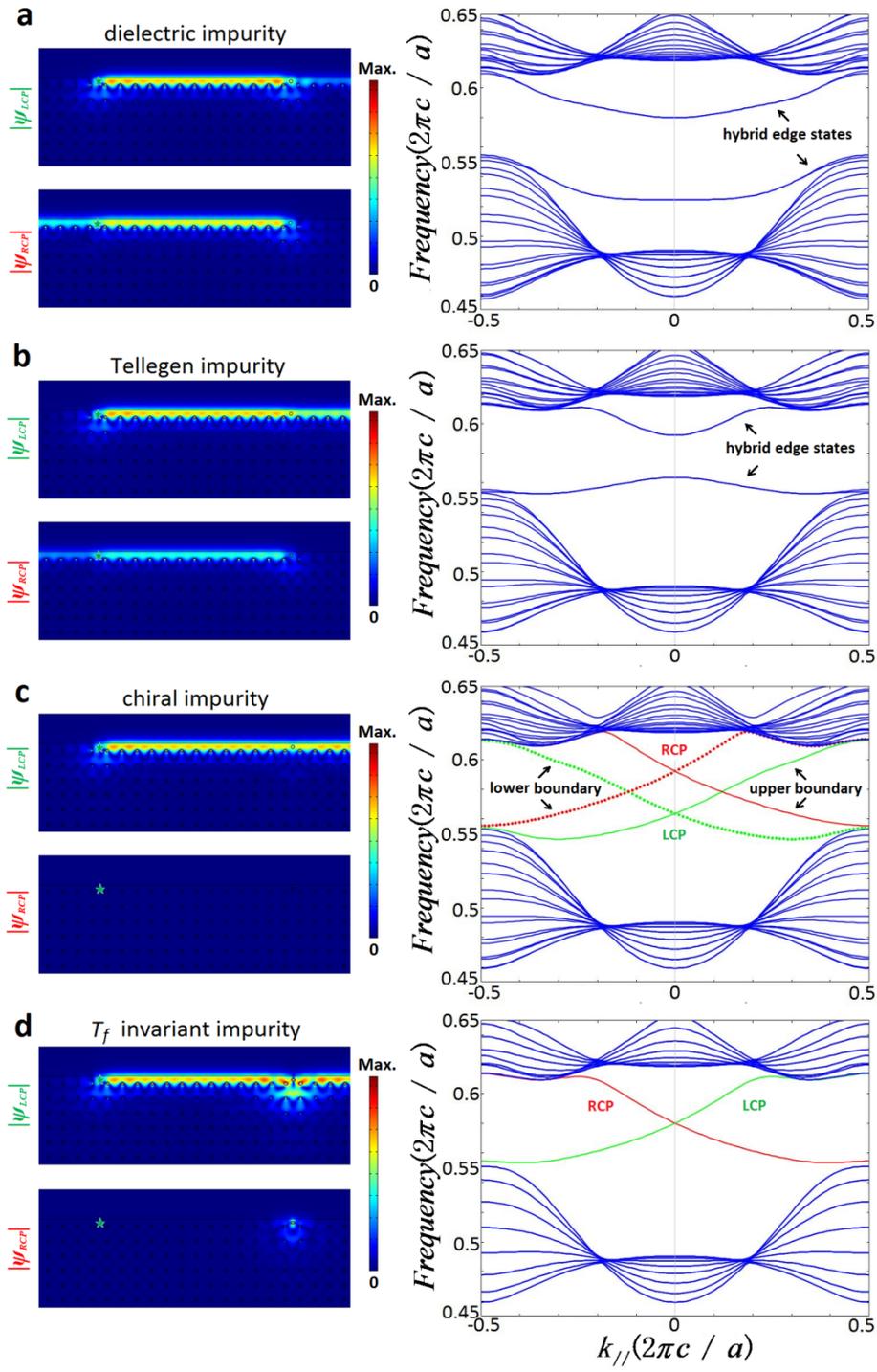

**Figure 4**